\documentclass[a4,usenatbib,fleqn]{mn2e}
\usepackage{mathptmx}
\usepackage{amsmath,amssymb,amsbsy}
\usepackage[pdftex]{graphicx}
\usepackage[backref=none]{hyperref}
\usepackage[usenames,dvipsnames]{color}
\usepackage{bm}
\usepackage{txfonts}
\DeclareSymbolFont{cmletters}{OML}{cmm}{m}{it}
\DeclareMathSymbol{v}{\mathalpha}{cmletters}{"76}

\definecolor{MyDarkBlue}{rgb}{0,0.1,0.7}
\hypersetup{pdfborder={0 0 0},colorlinks,breaklinks=true,
  urlcolor={blue},citecolor={MyDarkBlue},linkcolor={magenta}}

\newcommand{\apj}{ApJ}
\newcommand{\apjl}{ApJ}

\newcommand{\mnras}{MNRAS}
\newcommand{\nat}{Nature}

\newcommand{\aap}{A{\&}A}

\newcommand{\prc}{Phys. Rev. C}

\title[Neutron star spindown with time varying torques]{Neutron star dynamics under time dependent external torques}

\author[G\"{u}gercino\u{g}lu \& Alpar]{Erbil G\"{u}gercino\u{g}lu$^{1}$\thanks{Contact e-mail: \href{mailto:egugercinoglu@sabanciuniv.edu}{egugercinoglu@sabanciuniv.edu}} and M. Ali Alpar$^{1}$\thanks{Contact e-mail: \href{mailto:alpar@sabanciuniv.edu}{alpar@sabanciuniv.edu}}
\\
$^{1}$Faculty of Engineering and Natural Sciences, Sabanc{\i} University, Orhanl{\i}, Tuzla, 34956 Istanbul, Turkey}

\date{Accepted XXX. Received YYY; in original form ZZZ}

\pubyear{2017}

\begin{document}
\label{firstpage}
\pagerange{\pageref{firstpage}--\pageref{lastpage}}
\maketitle

\begin{abstract}
The two component model describes neutron star dynamics incorporating the response of the superfluid interior. Conventional solutions and applications involve constant external torques, as appropriate for radio pulsars on dynamical timescales. We present the general solution of two component dynamics under arbitrary time dependent external torques, with internal torques that are linear in the rotation rates, or with the extremely non-linear internal torques due to vortex creep. The two-component model incorporating the response of linear or nonlinear internal torques can now be applied not only to radio pulsars but also to  magnetars and to neutron stars in binary systems, with strong observed variability and noise in the spin-down or spin-up rates. Our results allow the extraction of the time dependent external torques from the observed spin-down (or spin-up) time series, $\dot{\Omega}(t)$. Applications are discussed.
\end{abstract}


%%%%%%%%%%%%%%%%%%%%%%%%%%%%%%%%%%%%%%%%%%%%%%%%%%

%%%%%%%%%%%%%%%%% BODY OF PAPER %%%%%%%%%%%%%%%%%%

% The MNRAS class isn't designed to include a table of contents, but for this document one is useful.
% I therefore have to do some kludging to make it work without masses of blank space.

\section{Introduction} \label{sec:intro}
Pulsar glitches are typically followed by exponential relaxation with timescales of days to weeks and nonlinear relaxation that 
can extend to the next glitch.  After the earliest observed glitches 
of the Crab and Vela pulsars \citet{baym69} pointed out that such 
slow relaxation indicates the presence of superfluid components of the 
neutron star, and proposed the two-component model for the dynamics of 
the neutron star crust coupled to a superfluid internal component: 
\begin{equation} 
I_{\rm c}\dot{\Omega}_{\rm c} + I_{\rm s}\dot{\Omega}_{\rm s} = N_{\rm ext}
\end{equation}
\label{eom}
\begin{equation}
\dot{\Omega}_{\rm s} = - \frac{\Omega_{\rm s} - \Omega_{\rm c}}{\tau_0} = - \frac{\omega}{\tau_0}
\label{dotsfluidlin}
\end{equation} 
where $I_{\rm c}$ and $I_{\rm s}$ are the moments of inertia and $\Omega_{\rm c}$ and $\Omega_{\rm s}$ 
are the rotation rates of the crust and the superfluid, respectively, and $\omega \equiv \Omega_{\rm s} - \Omega_{\rm c}$ 
is the rotational velocity lag between the two components. The internal torque that the superfluid exerts on the crust is
\begin{equation}
N_{\rm int} \equiv -I_{\rm s}\dot{\Omega}_{\rm s}.
\end{equation}  
This model explains the prompt exponential relaxation after a glitch. Note that the superfluid interior and solid crust in the neutron star bring about a fundamentally different structure to the rotational dynamics, depending on the lag $\omega \equiv \Omega_{\rm s} - \Omega_{\rm c}$ between the two components. By contrast, the rotational dynamics of a normal fluid star is described by the Navier-Stokes equation where the angular momentum transport by viscosity depends on the local gradients of the rotational velocity, $\nabla\Omega(r)$. In the two component models for neutron stars, the electrons that coexist with the neutron and proton superfluids in the interior are already coupled to the rigid rotation of the crust on timescales much shorter than the observed relaxation times after pulsar glitches. 

Typically several distinct components of relaxation with different 
relaxation times are observed after pulsars glitches 
\citep{alpar96,akbal16, erbil17}. After exponential relaxation is over 
the response to the glitch continues with a constant second 
derivative $\ddot{\Omega}_c$. Such power law behavior indicates the presence of nonlinear internal torques.
\citet{alpar84a} explained this nonlinear behavior in terms of vortex creep, the thermally activated motion of quantized vortices in the neutron star superfluid 
against inhomogeneities that can pin vortices.   

There is no evidence that the external torque is changing in pulsar glitches 
(this may not be the case for the exceptional glitch of PSR J 1119$-$6127  where the pulsar signature changes at the glitch possibly indicating a change in the external torque \citep{akbal15,archibald16}). 
Hence the observed glitch associated offsets $\Delta\dot{\Omega}_{\rm c}$ in the crust spindown rate 
and the subsequent relaxation give a measure of the ratio of moments of inertia 
of the crust and the superfluid components involved:
\begin{eqnarray}
\frac{\Delta\dot{\Omega}_{\rm c}}{\dot{\Omega}_{\rm c}}\cong \frac{I_{\rm s}}{I_{\rm c}} \nonumber. 
\end{eqnarray}
The fact that $\Delta\dot{\Omega}_{\rm c} / \dot{\Omega}_{\rm c} \cong I_{\rm s} / I_{\rm c} \cong I_{\rm s} / I \ll 1$ 
(where $I = I_{\rm c} + I_{\rm s}$ is the total moment of inertia) indicates that most of the neutron 
star's moment of inertia must be tightly coupled to the observed crust on timescales 
shorter than the postglitch relaxation timescales. The superfluid core that carries the bulk of the 
moment of inertia is indeed expected to couple to the outer crust and other normal 
(non-superfluid) components via the scattering of electrons from the neutron vortices, 
which carry a significant spontaneous magnetization due to induced currents of superconducting 
protons around the neutron vortex lines \citep{alpar84b}. Because $I_{\rm s}/I_{\rm c}$ is small one can treat 
the effect of each individual superfluid layer with its particular coupling to the crust separately 
and then add the effects of the different superfluid components on the 
dynamics of the crust. Taking into account the neutron effective mass in the inner crust lattice \citep{chamel12} 
increases the $I_{\rm s}$ values inferred from the observations somewhat, but it remains true that $I_{\rm s}/I_{\rm c}\lesssim 10\%$.

The two component model is generalized by replacing Equation (\ref{dotsfluidlin}) with  
\begin{equation}
\dot{\Omega}_{\rm s} = - f(\omega)
\label{dotsfluid}
\end{equation} 
to include nonlinear coupling. The original linear two component model is obtained when the coupling is linear in the lag, 
$f(\omega) = \omega / \tau_0$, while in the nonlinear regime of the vortex creep model 
$f(\omega) = C \exp (\omega/\varpi )$, $C$ and $\varpi$ being constants that depend on the temperature, pinning energy, average vortex density in the superfluid and steady state spindown rate \citep{alpar84a}. 

The equation for the evolution of the lag $\omega \equiv \Omega_{\rm s} - \Omega_{\rm c}$ is
\begin{equation}
\dot{\omega} = - \frac{I}{I_{\rm c}}f(\omega) - \frac{N_{\rm ext}}{I_{\rm c}}. 
\label{dotomega}
\end{equation} 
In the linear case we have 
\begin{align}
\dot{\omega}=&- \frac{I}{I_{\rm c}}\frac{\omega}{\tau_0} - \frac{N_{\rm ext}}{I_{\rm c}} \nonumber \\
=&-\frac{\omega}{\tau} - \frac{N_{\rm ext}}{I_{\rm c}},
\label{dotomegalin}
\end{align}
defining the relaxation time $\tau \equiv \frac{I_{\rm c}}{I}\tau_0$ of the system in terms of the coupling time $\tau_0$. 
 
The two component system has a steady state defined by
\begin{align}
\dot{\omega}&=0 \nonumber \\
\dot{\Omega}_{\rm s}&=\dot{\Omega}_{\rm c} = \frac{N_{\rm ext}}{I}. 
%% (8)
\end{align} 
The steady state value $\omega_{\infty}$ of the lag is determined by 
\begin{equation}
f(\omega_{\infty}) = - \frac{N_{\rm ext}}{I}.
%% (9)
\end{equation}
In the linear case this gives
\begin{equation}
\omega_{\infty}= - \frac{N_{\rm ext}}{I}\tau_0.
%% (10)
\end{equation}

In a glitch the crust's rotation rate increases by $\Delta\Omega_{\rm c}$. The superfluid rotation rate will decrease by $\delta\Omega_{\rm s}$ if
vortices unpinned in an avalanche in the glitch move through that superfluid region. The lag will be offset from the steady state by $\delta\omega = \Delta\Omega_{\rm c} + \delta\Omega_{\rm s}$. The subsequent post-glitch evolution back towards steady state is given by the solution of Equation (\ref{dotomega}) with the appropriate $f(\omega)$, i.e. with 
$f(\omega) = C \exp (\omega/\varpi)$ for nonlinear vortex creep and the solution of Equation (\ref{dotomegalin}) in the linear case. 
The postglitch relaxation of the superfluid rotation rate is then derived from Equation (\ref{dotsfluid}) (or Equation (\ref{dotsfluidlin}) in the linear case). Finally the postglitch relaxation of the observed crust rotation rate is calculated using Equation (\ref{eom}). 

This program has been developed for the effects of superfluid regions with both linear or nonlinear coupling to the crust, in particular detail for the Vela \citep{alpar84a, alpar93,akbal16} and Crab pulsar glitches \citep{alpar96}. Model fits to postglitch and interglitch timing behaviour yields information on the fractional moments of inertia and physical properties of the superfluid regions. The resulting understanding of internal torques allows one to predict the time of the next glitch with some accuracy. Furthermore, by subtracting the effect of internal torques one can extract the braking index $n$ that characterizes the external torque \citep{akbal16}. All applications of the two component model so far were made for radio pulsars, for which the external torque was taken to be constant, with good justification, since the secular timescale for changes in the pulsar external torque, the characteristic spindown time $\tau_{\rm sd} \equiv \Omega/ ( 2|\dot{\Omega}|)$, is much longer than observed 
dynamical timescales of postglitch relaxation or the observed intervals between subsequent glitches. Low level pulsar timing noise could be neglected without significant effect on the model fits.

By contrast, neutron stars in X-ray binaries, magnetars and transients exhibit strong variations in the observed spin-down or spin-up rates indicating variability in the external torque, including strong torque noise. The most comprehensive study on timing noise in pulsars by \citet{hobbs10} has revealed that for pulsars with characteristic ages $\tau_{\rm c}=10^{5}$ yr the dominant contribution to timing noise comes during the recovery from glitches. Some radio pulsars and magnetars show cyclic or oscillating components in their spin parameters \citep{lyne10,biryukov12,gao16} which can be interpreted as evidence for changing external torques. \citet{livingstone11} have reported that for the high magnetic field pulsar PSR J1846$-$0258 the level of the timing noise increased following its magnetar-like outburst. Magnetars do not have a stable spin down \citep{beloborodov17}. Variations in spindown rate are most pronounced following the outbursts and the glitches \citep{dib14}. High mass X-ray binaries also typically display strong variability and noise in their timing properties \citep{lvdk06}.

In this paper, we present the general solution of two component dynamics under arbitrary time dependent external torques, in conjunction with linear internal torques or with the extremely non-linear internal torques for non-linear vortex creep. Our work generalizes and extends the linear two component neutron star models developed by \citet{baym69} to explain postglitch evolution with constant external torques and by \citet{lamb78a,lamb78b,baykal91} to interpret timing noise in X-ray binaries in terms of transient epochs of accretion and sporadic unpinning events that give rise to time dependent torque variations and the nonlinear two component model of \citet{alpar84a}. In Section \ref{exttorconst} we summarize earlier results with linear or nonlinear internal torques and constant external torques. In Sections \ref{exttorlin} and \ref{exttornon} we present the general solutions for time dependent external torques with linear and non-linear internal torques, respectively. 
In Section \ref{application} we discuss applications with variable external torques, including timing noise, for neutron stars in binaries, magnetars and transients. Section \ref{dac} presents our conclusions. 
%%%%%%%%%%%%%%%%%%%%%%%%%%%%%%%%%%%%%%%%%%%%

\section{Two-Component Models with Constant External Torque} \label{exttorconst}

Linear internal torques, with $f(\omega)=\frac{\omega}{\tau_0}$ realize in many different physical contexts like mutual friction, drag forces between superfluid and normal matter components, or the linear regime of the vortex creep model. Equation (\ref{dotomegalin}) leads to the solution
\begin{equation}
\dot\Omega_{\rm c}= \dot\Omega_{\rm c}(0)-\frac{I_{\rm s}}{I}\frac{\delta\omega(0)}{\tau}e^{-t/\tau},
\label{dotOmegalin}
\end{equation}
under constant external torques. The linear response is simple exponential relaxation with an amplitude that is linear in the glitch induced perturbation $\delta\omega(0)$, and proportional to the moment of inertia fraction $I_{\rm s}/I $. For each specific model, $\tau$ depends on the physical parameters of the model.
In a region of superfluid where no glitch induced vortex motion takes place the offset $\delta\omega$ is simply the glitch $\Delta\Omega_{c}$ in the crust rotation rate.

In the vortex creep model a superfluid component with vortex pinning spins down by the thermally activated flow (creep) of vortices against pinning potentials. The spindown rate is 
\begin{equation}
\dot\Omega_{\rm s}=-\frac{4\Omega_{\rm s}v_0}{r}\exp{\left(-\frac{E_{\rm p}}{kT}\right)}\sinh{\left(\frac{\omega}{\varpi}\right)},
\label{dotsfluidcreep}
\end{equation}
where 
\begin{equation}
\varpi \equiv \frac{kT}{E_{\rm p}}\omega_{\rm cr}.
\end{equation}
Here $E_{\rm p}$ is the pinning energy, $T$ denotes the temperature, the distance $r$ of the vortex lines from the rotational axis is approximately equal to the neutron star radius $R$ in crustal layers and $v_{0}\approx 10^{7}$ cm/s is the microscopic vortex velocity around pinning centres \citep{alpar84a, erbil16}. Depending on the external torque and the values of $E_{\rm p}/kT$, Equation (\ref{dotsfluidcreep}) can indicate a linear regime when $\sinh{\left(\frac{\omega}{\varpi}\right)}\cong \frac{\omega}{\varpi}$ or a nonlinear regime, both of which seem to occur in neutron stars \citep{alpar89,erbil14}. In the linear regime, the response to a glitch is the linear response given in Equation (\ref{dotOmegalin}), with relaxation time 
\begin{equation}
\label{taulin}
\tau = \frac{I_{\rm c}}{I}\tau_{\rm l} \equiv \frac{I_{\rm c}}{I}\frac{kT }{E_{\rm p}} \frac{R \omega_{\rm cr}}{4 \Omega_{\rm s} v_{0}} \exp \left( \frac{E_{\rm p}}{kT} \right).
\end{equation}

In the nonlinear regime, the values of the parameters, in particular of $E_{\rm p}/kT$ are such that Equation (\ref{dotsfluidcreep}) requires $\sinh{(\omega/\varpi)}\gg 1$, so that $\sinh{(\omega/\varpi)} \simeq (1/2)\exp{(\omega/\varpi)}$, and 
\begin{equation}
f(\omega) = \frac{\varpi}{2\tau_{\rm l}}\exp(\omega/\varpi).
\end{equation}
Equations (\ref{eom}) and (\ref{dotsfluid}) then yield
\begin{equation}
\dot\Omega_{\rm c}= \frac{N_{\rm ext}}{I_{\rm c}} + \frac{I_{\rm s}}{I_{\rm c}}\frac{\varpi}{2\tau_{\rm l}}\exp(\omega/\varpi).
\label{ncreepeq}
\end{equation}
For a constant external torque $N_{\rm ext} = I \dot\Omega_{\infty}$ the solution for the observed crust spindown rate is
\begin{equation}
\dot\Omega_{\rm c}(t)=\frac{I}{I_{\rm c}}\dot\Omega_{\infty}-\frac{I_{\rm s}}{I_{\rm c}}\dot\Omega_{\infty} \left[1-\frac{1}{1+\left[\exp{\left(\frac{t_0}{\tau_{\rm nl}}\right)}-1\right]\exp{\left(-\frac{t}{\tau_{\rm nl}}\frac{I}{I_{\rm c}}\right)}}\right],
\label{ncreep}
\end{equation}
with a nonlinear creep relaxation time
\begin{equation}
\tau_{\rm nl}\equiv \frac{kT}{E_{\rm p}}\frac{I\omega_{\rm cr}}{N_{\rm ext}} =\frac{kT}{E_{\rm p}}\frac{\omega_{\rm cr}}{\vert\dot{\Omega}\vert_{\infty}},
\label{taun}
\end{equation}
and recoupling (waiting) timescale
\begin{equation}
t_0 \equiv \frac{I\delta\omega}{N_{\rm ext}}=\frac{\delta\omega}{\vert \dot\Omega\vert_{\infty}}.
\label{t0}
\end{equation}
When the external torque is taken to be constant during a glitch angular momentum conservation leads to $\delta \Omega_{\rm s} =(I_{\rm c}/I_{\rm s})\Delta \Omega_{\rm c}\gg\Delta\Omega_{\rm c}$, and $\delta\omega\cong\delta\Omega_{\rm s}$. Nonlinear creep regions are responsible for glitches through vortex unpinning avalanches and creep restarts after a waiting time $t_{0}\cong\delta\Omega_{\rm s}/\vert\dot{\Omega}\vert_{\infty}$. The approximately constant second derivative $\ddot{\Omega}_{\rm c}$ interglitch timing behaviour observed in the Vela and other pulsars \citep{akbal16,yu13} corresponds to a uniform density of vortices unpinned at the glitch leading to a range of waiting times throughout the superfluid regions. 

Two component models with linear or nonlinear internal torques have so far been applied to post-glitch or inter-glitch timing behaviour of radio pulsars \citep{alpar93,alpar96,akbal15,akbal16,erbil17}, where the external torque, with a secular (characteristic) timescale $\tau_{\rm c} \equiv \Omega_{\rm c}/2|\dot{\Omega}_{\rm c}| \sim 10^3 - 10^6 $ yr is constant for timescales of observed postglitch relaxation.  
%%%%%%%%%%%%%%%%%%%%%%%%%%%%%%%%%%%%%%%%%%%%

\section{Linear Two-Component Model with a Time Dependent External Torque} \label{exttorlin}

Neutron stars in (especially high mass) X-ray binaries and magnetars display strongly variable timing behaviour on observational timescales. The postglitch response with a linear internal torque and a time varying external torque is described by 
\begin{equation}
\dot{\omega} = - \frac{\omega}{\tau} - \frac{N_{\rm ext}(t)}{I_{\rm c}}.
\label{dotomegalt}
\end{equation}
The relaxation time $\tau$ is given by the physical processes of the internal torque. On multiplying both sides with an integration factor $e^{t/\tau}$ we obtain
\begin{equation}
\frac{d}{dt}\left[\omega e^{t/\tau}\right]=-\frac{N_{\rm ext}(t)}{I_{\rm c}}e^{t/\tau},
\end{equation}
which gives
\begin{equation}
\omega(t)=e^{-t/\tau}\left[\omega(0)-\frac{1}{I_{\rm c}}{\int_{0}^{t}e^{t'/\tau}}N_{\rm ext}(t')dt'\right].
\label{omegal}
\end{equation}
Equations (\ref{eom}), (\ref{dotomegalt}), and (\ref{omegal}) yield 
\begin{equation}
\dot\Omega_{\rm c}(t)=\frac{N_{\rm ext}(t)}{I_{\rm c}}+\frac{I_{\rm s}}{I}\left(\frac{e^{(-t/\tau)}}{\tau}\left[\omega(0)-\frac{1}{I_{\rm c}}\int_{0}^{t} e^{(t'/\tau)}N_{\rm ext}(t') d t'\right]\right).
\label{extlin}
\end{equation}

\section{Nonlinear Vortex Creep Model with a Time Dependent External Torque} \label{exttornon}

For nonlinear vortex creep Equations (\ref{dotomega}) and (\ref{dotsfluidcreep}) lead to
\begin{equation}
\dot\omega=-\frac{I \varpi}{2I_{\rm c}\tau_{\rm l}}e^{\omega/\varpi}-\frac{N_{\rm ext}(t)}{I_{\rm c}},
\label{dotomeganl}
\end{equation}
with a time dependent external torque. Defining $y\equiv \exp(-\omega/\varpi)$, Equation (\ref{dotomeganl}) becomes a Bernoulli type equation:
\begin{equation}
\frac{dy}{dt}-\frac{N_{\rm ext}(t)}{I_{\rm c}\varpi}y-\frac{I}{2I_{\rm c}\tau_{\rm l}}=0.
\end{equation}
This equation has an integration factor $\exp\left(-\frac{X(t)}{I_{\rm c}\varpi}\right)$ where 
\begin{equation}
\frac{dX(t)}{dt}=N_{\rm ext}(t) \Longleftrightarrow X(t)=\int_{0}^{t}N_{\rm ext}(t') d t'.
\end{equation}
The angular velocity lag $\omega$ displays an exponential dependence on the glitch induced changes in $\omega(0)$ and on $X(t)$, the cumulative angular momentum transfer by the external torque:
\begin{align}
e^{-(\omega/\varpi)}&=e^{-\omega(0)/\varpi}\exp\left(\frac{X(t)}{I_{\rm c}\varpi}\right)\nonumber \\
 &+ \exp\left(\frac{X(t)}{I_{\rm c}\varpi}\right)\int_{0}^{t}\frac{I}{2I_{\rm c}\tau_{\rm l}}\exp\left(-\frac{X(t')}{I_{\rm c}\varpi}\right)dt'.
\label{omeganl}
\end{align}
The most general response of nonlinear creep to a glitch in the presence of a time dependent external torque is obtained from Equations (\ref{eom}), (\ref{dotomeganl}), and (\ref{omeganl}):
\begin{align}
\dot\Omega_{\rm c}(t)=&\frac{N_{\rm ext}(t)}{I_{\rm c}}\nonumber \\
&+\frac{I_{\rm s}}{I_{\rm c}}\frac{\varpi}{2\tau_{\rm l}}\left[\frac{1}{\exp{\left(\frac{\frac{X(t)}{I_{\rm c}}-\omega(0)}{\varpi}\right)}+\exp{\left(\frac{X(t)}{I_{\rm c}\varpi}\right)}\int_{0}^{t}\frac{I}{2I_{\rm c}\tau_{\rm l}}\exp{\left(-\frac{X(t')}{I_{\rm c}\varpi}\right)}dt'} \right].
\label{extnlin}
\end{align}

By introducing the external torque averaged over the period of time $t$ as $<N_{\rm ext}(t)>= X(t)/t$ the nonlinear creep timescale and the waiting time can be expressed by
\begin{equation}
\tau_{\rm nl}(t) \equiv \frac{kT}{E_{\rm p}}\frac{I\omega_{\rm cr}}{<N_{\rm ext}(t)>},
\end{equation}
and
\begin{equation}
t_{0}(t) \equiv \frac{I\delta \omega(0) }{<N_{\rm ext}(t)>},
\end{equation}
for the case of a time dependent external torque.
In contrast to the case of constant external torque, Equations (\ref{taun}) and (\ref{t0}), the nonlinear creep time scales themselves are now time dependent, involving the running time average of the external torque.

\section{Applications } \label{application}

The neutron star core superfluid contains most of the moment of inertia of the star so that $I_{\rm s, core}/I \sim 1$ . This makes the `direct' and integrated response terms in Equations (\ref{extlin}) and (\ref{extnlin}) of comparable magnitude. The internal torques coupling the core superfluid  to the crust are expected to display linear response. The expected coupling times are $(10-200)P$ in the case of electron scattering from spontaneously magnetized neutron vortex lines \citep{alpar84b,sidery09} and therefore do not contribute to the observed response except possibly in the longest period neutron stars. When flux tube--vortex line interactions are taken into account the coupling times become much shorter \citep{sidery09,erbil16}. Thus, core superfluid effects will not be observable. Indeed, all changes in $\dot\Omega_{\rm c}$ remain within $\frac{\Delta\dot\Omega_{\rm c}}{\dot\Omega_{\rm c}}\lesssim 10^{-2}-10^{-1}$. This indicates that only the crustal superfluid and toroidal field region in the outer core can take part in observable phenomena \citep{erbil14}. 

For the crust and outer core toroidal field region superfluids the ratio $I_ {\rm s} / I$ is of order 0.1 or less as deduced from the glitch data with constant external torque \citep{erbil14,erbil17}. The direct term $N_{\rm ext}(t)/I_{\rm c}$ therefore dominates in the solution Equations (\ref{extlin}) and (\ref{extnlin}). This fact allows us to follow a simple strategy: model fits can start by fitting the direct term to the data first to extract $N_{\rm ext}(t)=I\dot\Omega_{\rm c}(t)$. The model can then be checked by comparing the integrated (second) term to the residuals of the first fit. For the linear response case we have exact analytical expressions. The nonlinear coupling case is more complicated, and has to be handled numerically. In any case, the same strategy applies. 

We will consider three particular cases for time dependent external torques; (i) an exponentially decaying external torque, (ii) power law time dependence and (iii) timing noise. 

Evaluating Equation (\ref{extlin}) for an exponentially decaying torque with a time scale $\tau_{\rm d}$ added to the preglitch external torque $N_0$ yields
\begin{equation}
N_{\rm ext}(t)=N_0 + \delta N e^{-t/\tau_{\rm d}}=I\dot\Omega_{\infty} + \delta N e^{-t/\tau_{\rm d}}.
\end{equation}

For linear internal torques Equation (\ref{extlin}) gives the solution
\begin{align}
\dot\Omega_{\rm c}(t)&=\dot\Omega_{\infty}+\frac{\delta N}{I_{\rm c}}e^{-t/\tau_{\rm d}}\left[1-\frac{I_{\rm s}}{I}\frac{\tau_{\rm d}}{\tau_{\rm d}-\tau}\right]\nonumber \\ 
&+\frac{I_{\rm s}}{I}\left[e^{-t/\tau}\left(\frac{\omega(0)}{\tau}+\frac{I}{I_{\rm c}}\dot\Omega_{\infty}+\frac{\delta N}{I_{\rm c}}\frac{\tau_{\rm d}}{\tau_{\rm d}-\tau}\right)\right].
\end{align}

For nonlinear creep the solution given in Equation (\ref{extnlin}) depends on 
the integrated angular momentum transfer $X(t) $ which is 
\begin{equation}
X(t)=N_{0} t + \delta N \tau_{\rm d} [ 1 - e^{-t/\tau_{\rm d}}], 
\end{equation}
for the exponentially decaying  external torque. Substituting this in Equation (\ref{extnlin}) gives the solution. 

Second, we evaluate Equation (\ref{extlin}) for a power law  torque with index $\alpha$ added to the preglitch torque:
\begin{equation}
N_{\rm ext}(t)=N_0 + \frac{\delta N t_0^{\alpha}}{(t+t_0)^{\alpha}}=I\dot\Omega_{\infty} + \frac{\delta N t_0^{\alpha}}{(t+t_0)^{\alpha}}.
\end{equation}
Equation (\ref{extlin}) gives
\footnotesize
\begin{align}
\lefteqn{\dot\Omega_{\rm c}(t)=\dot\Omega_{\infty}+\frac{\delta N}{I_{\rm c}}\frac{t_0^{\alpha}}{(t+t_0)^{\alpha}}\left[1-\frac{I_{\rm s}}{I}\frac{(t+t_0)}{(1-\alpha)\tau}\right]{}}\nonumber \\
&&{}+\frac{I_{\rm s}}{I}\left[e^{-t/\tau}\left(\frac{\omega(0)}{\tau}+\frac{I}{I_{\rm c}}\dot\Omega_{\infty}+\frac{\delta N}{I_{\rm c}}\frac{t_0}{(1-\alpha)\tau}+\frac{I_{\rm s}}{I}\frac{\delta N}{I_{\rm c}}\frac{t_0^{\alpha}}{(1-\alpha)\tau^{2}}\int_{0}^{t} \frac{e^{t'/\tau}dt'}{(t+t_0)^{\alpha-1}}\right)\right],
\end{align}
\normalsize
for linear internal torques. The integrated angular momentum transfer $X(t) $ is 
\begin{equation}
X(t)=N_0 t + \frac{\delta N t_0}{\alpha - 1}\left[ 1 - \frac{t_0^{\alpha - 1}}{(t+t_0)^{\alpha - 1}}\right],
\end{equation}
defining the response with nonlinear internal torques through Equation (\ref{extnlin}). 

To illustrate noise processes, we choose white torque noise in the linear response regime. Noise processes are best described in terms of power spectra. White noise corresponds to spikes occurring at random times in the data, and is expressed in terms of Dirac delta functions as
\begin{equation}
N_{\rm ext}=\sum_{\rm i}\alpha_{\rm i}\delta(t-t_{\rm i}),
\label{wnoise}
\end{equation}
where 
$\alpha_{\rm i}$ are amplitudes of torque variations. For linear coupling Equations (\ref{extlin}) and (\ref{wnoise}) lead to the power spectrum
\begin{align}
P(f)=\frac{1}{\sqrt{2\pi}}\left[\frac{2<\alpha^{2}>}{I_{\rm c}^{2}}+\left(\frac{I_{\rm s}/I}{1+(2\pi\tau f)^{2}}\right)\left(\frac{<\alpha><\omega>}{I_{\rm c}}-\frac{<\alpha^{2}>}{I_{\rm c}^{2}}\right)\right] \nonumber \\
+\frac{1}{\sqrt{2\pi}}\left[\left(\frac{(I_{\rm s}/I)^{2}}{1+(2\pi\tau f)^{2}}\right)\left(\frac{2<\alpha^{2}>}{I_{\rm c}^{2}}+2<\omega^{2}>-\frac{<\alpha><\omega>}{I_{\rm c}}\right) \right],
\label{powerspec}
\end{align}
where $<\alpha>$ is the mean of external torque variation amplitude and $<\omega>$ denotes the mean value of the angular velocity lag. The solution (\ref{powerspec}) can be compared with the timing data. To lowest order we find the model power spectrum, a constant $P(f)$ for white noise. To order $I_{\rm s}/I$ we find the power spectrum of the integrated process, which is flat at low $f$ and a random walk spectrum, $P(f)\propto f^{-2}$, at high frequencies $f\gg \tau^{-1}$. In line with our strategy, the first fit will give the strength of the noise process, the term proportional to $<\alpha^{2}>$. The residuals, to order $I_{\rm s}/I$ will yield information on the other parameters $<\alpha><\omega>$ and $<\omega^{2}>$. The term proportional to $(I_{\rm s}/I)^{2}$ can be neglected to a good approximation. For other noise processes, we expect to find the power spectrum of the torque noise from the lowest order fit to the observed power spectrum of $\dot\Omega_{\rm c}$, and as far as the resolution of the timing data allows, the residuals to order $I_{\rm s}/I$ would contain the power spectrum of the corresponding integrated process. Comparison of power spectra before and after a glitch can yield information on the glitch-noise correlation.

%%%%%%%%%%%%%%%%%%%%%%%%%%%%%%%%%%%%%%
\section{Conclusions} \label{dac}

We have presented, for the first time, the general solution of the temporal behaviour of a two component neutron star under time dependent torques. Both linear and nonlinear internal torques are taken into consideration. Analytical expressions are obtained with arbitrary time dependent external torques. We have explored the specific examples of exponentially decaying and power law  external torques as well as white torque noise. To the lowest approximation, the observed $\dot\Omega_{\rm c}(t)$ reflects the external torque. Contributions of the integrated superfluid response are of the order of the moment of inertia fraction of the superfluid component, $I_{\rm s}/I \lesssim 10^{-1}$. The superfluid parameters can be extracted from the data after removing the external torque term $N_{\rm ext}(t)/I_{\rm c}$ as the lowest order fit to $\dot\Omega_{\rm c}(t)$. The residuals reflect the contribution in proportion to $I_{\rm s}/I$ of the integrated response of internal superfluid torque and its coupling to the variation in the external torque. The amount of superfluid participating in the event, the strength of the superfluid-crust coupling and parameters of the superfluid relaxation process can be inferred from the observations. In the linear response case the creep relaxation timescale decouples from the external torque so that it is a constant solely dependent on superfluid-normal matter interaction parameters in the inner crust. Linear response allows for analytical solutions for spindown rate. Numerical calculations are needed in the case of nonlinear coupling. Nonlinear relaxation timescales involve mean spin-down rates and thus change along with the data span used. Information on the superfluid components, and therefore on the neutron star structure can thus be extracted for neutron stars under variable external torques. The model can be tested for consistency as the residuals of order $I_{\rm s}/I$ are correlated with the lowest order term $N_{\rm ext}(t)/I_{\rm c}$ in specified ways. The form of the external torque is determined from the preglitch epoch and picked up from the post event data. Our results are applicable to noisy radio pulsars, to the timing behaviour of magnetars and high mass X-ray binaries.   

\section*{Acknowledgements}
% Entry for the table of contents, for this guide only
\addcontentsline{toc}{section}{Acknowledgements}

This work was supported by the Scientific and Technological Research Council of Turkey
(T\"{U}B\.{I}TAK) under the grant 113F354. M.A.A. is a member of the Science Academy
(Bilim Akademisi), Turkey. We thank the referee for carefully reading the manuscript and for clarifying comments.
%%%%%%%%%%%%%%%%%%%%%%%%%%%%%%%%%%%%%%%%%%%%%%%%%%



%\bibliographystyle{mnras}


% Don't change these lines
\bsp	% typesetting comment
\label{lastpage}
\end{document}